\documentclass[aps,prb,twocolumn,superscriptaddress,10pt]{revtex4-2} 
\usepackage{graphicx,graphics} 
\usepackage{epstopdf} 
\usepackage{dcolumn}
\usepackage{amsmath,amssymb,amsfonts} 
\usepackage{latexsym,verbatim} 
\usepackage{bm} 
\usepackage{bbold} 
\usepackage{xcolor} 
\usepackage{ulem} 
\usepackage[breaklinks=false,colorlinks,citecolor=blue,linkcolor=blue,urlcolor=blue]{hyperref} 
\usepackage[abs]{overpic} 
\usepackage{textcomp} 
\usepackage{mathtools} 
\usepackage{braket} 
\usepackage{soul} 
\usepackage{lipsum}
\usepackage[export]{adjustbox}
\usepackage{gensymb}
\usepackage[T1]{fontenc}
\usepackage{multirow}
\usepackage{cancel}

\begin{document}

\title{Hole-spin qubits in germanium beyond the single-particle regime}

\author{Andrea Secchi}
\email{andrea.secchi@nano.cnr.it}
\affiliation{Centro S3, CNR-Istituto Nanoscienze, 41125 Modena, Italy}

\author{Gaia Forghieri}
\altaffiliation{Present address: Dipartimento di Fisica, Università di Milano, I-20133 Milan, Italy}
\affiliation{Centro S3, CNR-Istituto Nanoscienze, 41125 Modena, Italy}
\affiliation{Dipartimento di Scienze Fisiche, Informatiche e Matematiche,
Università degli Studi di Modena e Reggio Emilia, Via G. Campi 213/a, 41125 Modena, Italy}

\author{Paolo Bordone}
\affiliation{Centro S3, CNR-Istituto Nanoscienze, 41125 Modena, Italy}
\affiliation{Dipartimento di Scienze Fisiche, Informatiche e Matematiche,
Università degli Studi di Modena e Reggio Emilia, Via G. Campi 213/a, 41125 Modena, Italy}

\author{Daniel Loss}
\affiliation{Department of Physics, University of Basel, Klingelbergstrasse 82, CH-4056 Basel, Switzerland}
\affiliation{Physics Department, King Fahd University of Petroleum and Minerals, 31261, Dhahran, Saudi Arabia} 
\affiliation{Quantum Center, KFUPM, Dhahran, Saudi Arabia} 
\affiliation{RDIA Chair in Quantum Computing}

\author{Stefano Bosco}
\affiliation{QuTech and Kavli Institute of Nanoscience, Delft University of Technology, Delft, The Netherlands}

\author{Filippo Troiani}
\affiliation{Centro S3, CNR-Istituto Nanoscienze, 41125 Modena, Italy}

\date{\today}

\begin{abstract}
The intense simulation efforts on hole-spin qubits in germanium have so far focused primarily on singly occupied quantum dots. Here, we theoretically investigate three-hole qubits in germanium and demonstrate that their performance can rival that of single-hole qubits in both strained and unstrained systems. In particular, we find that --- in the widely used quasi-circular geometry --- a three-hole qubit encoding can yield enhancements of the Rabi frequencies of up to two orders of magnitude and a large advantage also in terms of quality factors.
\end{abstract}

\maketitle

\begin{figure}[t]
    \begin{center}
    \includegraphics[width=0.4\textwidth]{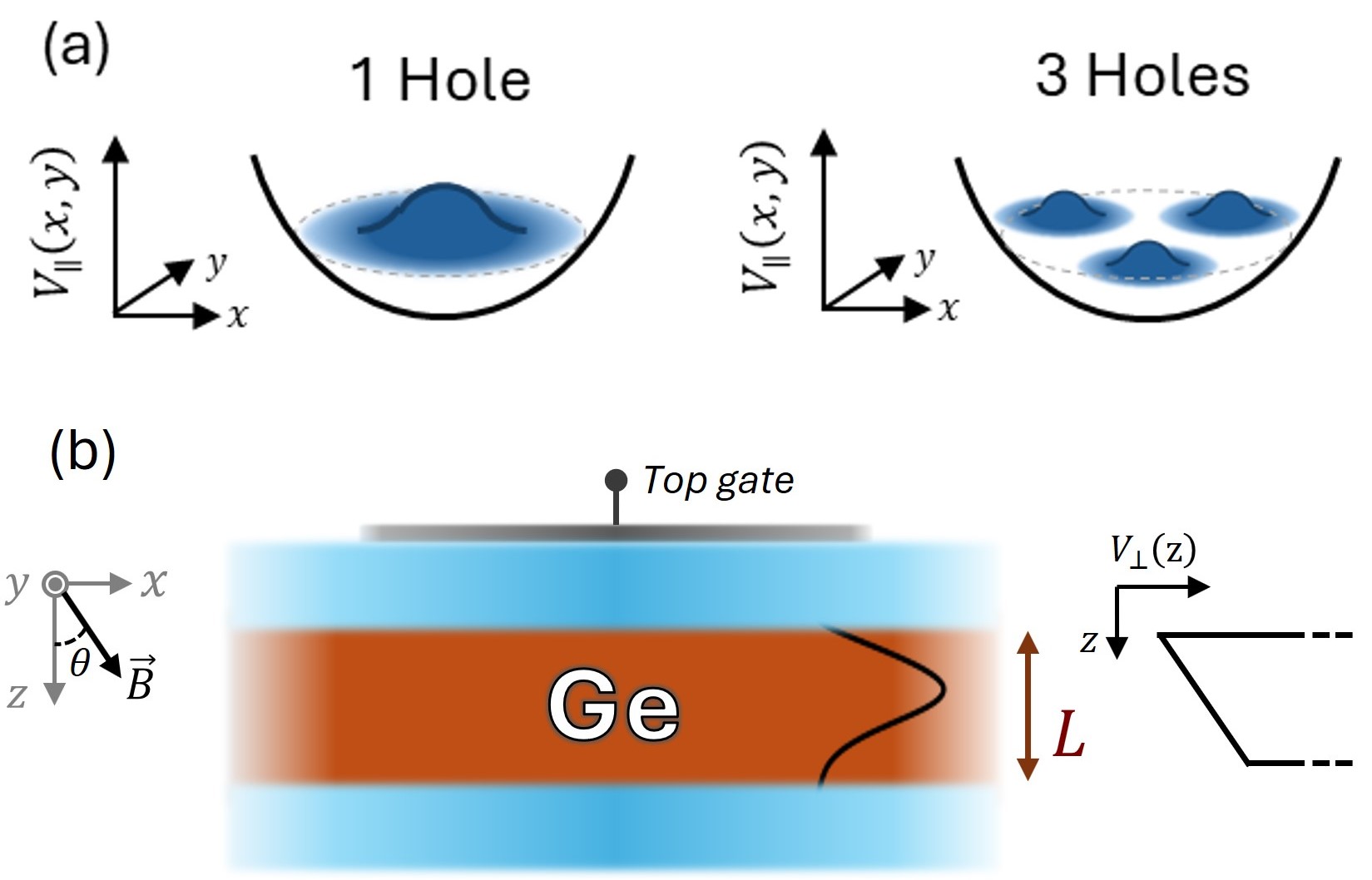}
    \caption{(a) Pictorial representation of a single-hole state (left) and a three-hole state (right) in a QD elongated along the $x$ direction. (b) Schematic view of a MOS device, where the Ge channel has a width $L$ along the $z$ direction. The bias applied to the metal top gate determines the in-plane confining potential $V_\parallel$ and the linear (in $z$) contribution in the out-of-plane confining potential $V_\perp$.} 
    \label{fig0}
    \end{center}
\end{figure}

\begin{figure}[t]
    \centering 
    \includegraphics[width=0.4\textwidth]{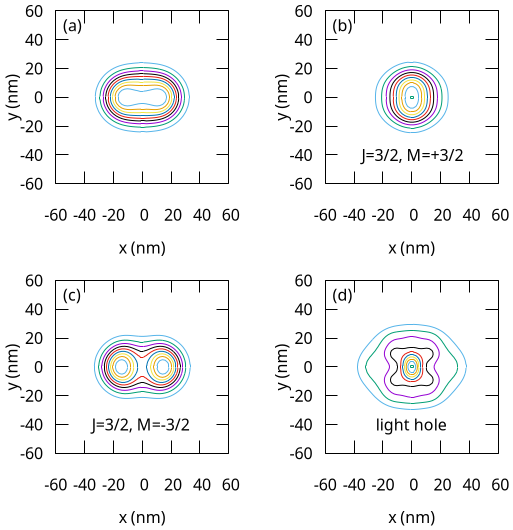}
    \caption{In-plane particle distribution $N_z(x,y)$ for the ground state $|0^{(3)}\rangle$ of the unstrained three-hole system, obtained by integrating the particle density $n(\boldsymbol{r})$ along the $z$ direction. The four panels display: (a) the function $N_z(x,y)$; (b,c) the heavy-hole contributions; (d) the sum of the two light-hole contributions. All the plotted functions are normalized to 1, the isolines correspond to values that are multiples of 0.1.  The values of the parameters are: $\hbar\omega_x=5\,$meV, $\hbar\omega_y=6\,$meV, $B=0.05\,$T and $\theta=0$ (parallel field orientation).} 
    \label{fig1}
\end{figure}

\section{Introduction}
Spin qubits in semiconductor quantum dots (QDs) are promising competitors for the implementation of quantum computers \cite{loss1998, burkard2023, stano2022, vandersypen2017, fang2023, scappucci2021, Maurand2016, Jirovec2021, Steinacker2024, Liles2024, camenzind2022, Geyer2024, Neyens2024, xue2021, zwerver2022}. In particular, the low natural abundance of magnetic nuclei, combined with the possibility of isotopic purification, grants long coherence times for spins in silicon (Si) and germanium (Ge)  \cite{stano2022,burkard2023,kobayashi2021,piot2022,huang2024,philips2022,Zhang2025,hendrickx2021,Wang2024Sc,George2025}. In these systems, high fidelities have been achieved in all fundamental computational tasks, from one- and two-qubit gates \cite{noiri2022,yoneda2018,xue2022,lawrie2021,mills2022,Weinstein2023} to initialization and readout \cite{mills2022-2,Kobayashi2023,takeda2024}. Besides, the advanced fabrication techniques and integration circuitry provide excellent scalability prospects and allow a high degree of control even at relatively high temperatures \cite{demichielis2023,zwerver2022,saraiva2022,Bellentani21a,camenzind2022,xue2021,petit2022,huang2024}. 

In contrast to electrons, holes in Ge have no valley degeneracy and are characterized by strong spin-orbit coupling (SOC) which allows for all-electric manipulation \cite{Bulaev2005,Bulaev2007,hendrickx2021,bosco2021-2,bosco2021-3,liles2021,michal2021, John2024, Bosco2023, carballido2024, Kloeffel2018, Adelsberger2022, AbadilloUriel2023}. In addition, due to their $p$-orbital character, they are less susceptible than electrons to the hyperfine interaction \cite{Fischer2008,scappucci2021,bosco2021,Cvitkovich2024}. Finally, 
the characteristic properties of the hole-spin qubit, including the $g$ tensor and the Rabi frequencies, are electrically tunable \cite{Bellentani21a,Venitucci18}. This results from the susceptibility to the electrostatic environment of the mixing between
heavy-hole (HH), light-hole (LH), and split-off (SO) bands \cite{Secchi21a, Stano2025, ares2013,forghieri2023, bosco2021-2, malkoc2022, Geyer2024, Jirovec2021, Wang2024}. 

The vast majority of the literature on hole-spin qubits focuses on single-particle encodings. Multiple holes, localized in singly-occupied QDs, have also been considered for alternative qubit encodings \cite{Liles2024,hetenyi2022,mutter2021,jirovec2022,fernandez2022, SaezMollejo2025, Zhang2025, Bosco24}. On the other hand, an odd number of holes in a {\it single} QD enables qubit encodings and manipulation schemes analogous to those implemented in the single-particle case \cite{lawrie2020,Voisin2016,Liu2022}. Such a multi-hole single-dot qubit has a twofold interest. First, from an experimental point of view, it relaxes the stringent requirement of achieving the single-occupation regime. Second, and most importantly, it might result in larger and more local couplings to the control gates and, ultimately, in improved figures of merit \cite{John25a}. However, despite their past and recent occurrence in experiments \cite{camenzind2022, Geyer2024, Bosco2023, carballido2024, John25a, DePalma2024}, few-particle effects in semiconductor hole-spin qubits remain relatively unexplored at the theoretical level. 

In this work, we provide a theoretical study of multi-hole qubits in Ge QDs, implemented in unstrained MOS-like devices and in Ge$/$Si$_{1-x}$Ge$_x$ heterostructures (Fig. \ref{fig0}). Focusing on the case of three holes, we show that the main figures of merit are always at least comparable to those obtained for single holes, and significantly improved in specific regimes. More specifically, the Rabi frequency of the three-hole qubit (THQ) is shown to exceed that of a single-hole qubit (SHQ) by up to two orders of magnitude in quasi-circular QDs. This occurs without a significant modification of the $g$ factor or loss of the qubit coherence under charge noise. As the comparison of the fully interacting THQ with its noninteracting counterpart shows, this gain is caused, for the largest part, by the occupation of the excited orbitals dictated by the Pauli principle, the hole-hole interaction giving a quantitatively smaller (albeit far from negligible) contribution in that regime. Strain changes the absolute values of the manipulation frequencies, but does not substantially affect the relative performances of the single- and multi-hole qubits. 

\section{Quantum-dot model}
Single-hole states are derived through the numerical diagonalization of the 6-band $\boldsymbol{k}\cdot\boldsymbol{p}$ Hamiltonian resulting from the Luttinger-Kohn (LK) envelope-function approach \cite{Luttinger55, Voon09,Secchi21a,Terrazos2021}. The 6 bands needed to describe single-hole states in Ge close to the top of the valence band form a $J=3/2$ quadruplet and a $J=1/2$ doublet, with $J$ being the total atomic angular momentum associated with the Bloch states at the $\Gamma$ point. These two multiplets are split at $\Gamma$ by spin-orbit coupling (SOC), which induces a gap $\Delta_{\rm SO}=290$ meV. The $J=3/2$ quadruplet, degenerate at $\Gamma$, is split into the HH ($|M| = 3/2$) and the LH ($|M|=1/2$) doublets at finite $k$. The band structure around the $\Gamma$ point is determined by the LK parameters, which, for Ge, are: $\gamma_1=13.38$, $\gamma_2=4.24$ and $\gamma_3=5.69$. The reference frame, specified by the unit vectors $\boldsymbol{u}_x$, $\boldsymbol{u}_y$, and  $\boldsymbol{u}_z$, is defined with respect to the crystallographic axes as follows: $\boldsymbol{u}_x \parallel [110]$, $\boldsymbol{u}_y  \parallel [\bar{1}10]$, $\boldsymbol{u}_z \parallel [001]$.

\begin{table}[b]
\centering
\renewcommand{\arraystretch}{1.5}
\begin{tabular}{|c|cc|cc|}
\hline
QD1 & $g^{(1)}_\theta$ & $g^{(3)}_\theta$ & $f^{(3)}_{R,x}$ & $f^{(3)}_{R,y}$ \\
\hline
$\theta=0^\circ$ & 8.67 & 7.99 (6.25) & 46.8 (239) & 9.24 (27.3) \\
$\theta=90^\circ$ & 0.121 & 0.0848 (0.104) & 7.29 (3.93) & 2.07 (0.0775)\\
\hline
\hline
QD2 & $g^{(1)}_\theta$ & $g^{(3)}_\theta$ & $f^{(3)}_{R,x}$ & $f^{(3)}_{R,y}$ \\
\hline
$\theta=0^\circ$ & 8.00 & 4.88 (4.34) & 394 (322) & 313 (175) \\
$\theta=90^\circ$ & 0.114 & 0.220 (0.321) & 27.1 (23.9) & 11.3 (2.04) \\ 
\hline
\end{tabular}
\renewcommand{\arraystretch}{1}
    \caption{Values of the $g$ factors and of the Rabi frequencies (expressed in MHz) for the unstrained single- and three-hole qubits, for $\hbar \omega_y = 6\,$meV, $B = 0.05$ T, and $\left| \delta \boldsymbol{E}_{\rm R} \right| = 1$ mV$/$nm. The value of $\hbar \omega_x$ is 3 and $5.5\,$meV for QD1 and QD2, respectively. In parentheses we report the Rabi frequencies of the noninteracting THQ.}
    \label{TableI}
\end{table}

The confining potential that defines the dot model varies smoothly with respect to the unit-cell scale, and is specifically given by
\begin{align}
V_{\rm conf}(\boldsymbol{r}) = \frac{1}{2} \frac{m_0}{\gamma_1} \left( \omega_x^2 x^2 + \omega_y^2 y^2 \right) + V_{\perp}(z) \,.
\end{align}
Here, $m_0$ is the free-electron mass, while $\hbar \omega_y$ (set to 6 meV) and $\hbar \omega_x$ (from 3 to 5.5 meV) are the characteristic energy scales associated with the anisotropic harmonic confinement in the $xy$ plane. The out-of-plane confinement defines a quantum well, resulting from the band offset between Ge and the barrier material: 
\begin{align}
    V_{\perp}(z) \!=\! W\!+\!\frac{W}{2}\sum_{\xi=\pm 1}\tanh\left[ \xi \frac{z}{s} \!-\!  \frac{(1+ \xi)}{2} \frac{L}{s} \right] \!+\! e_{\perp} z\, .
\end{align}
Here, $L = 10\,$nm is the well width, $W$ is the barrier height, and $s= 0.02\,$nm a smoothing factor. The dc electric field $ e_{\perp} = 1\,$meV$/$nm, possibly associated to a top gate voltage, is attractive for holes. 

The coupling to the magnetic field is introduced into the single-hole Hamiltonian through the Zeeman-Bloch term and the Peierls substitution \cite{Venitucci18, Bellentani21a}, with the electromagnetic vector potential expressed in the Coulomb gauge. 
The magnetic field is given in polar form, $\boldsymbol{B} \equiv B\, ( \sin\theta\,\cos\phi, \, \sin\theta\,\sin\phi, \, \cos\theta ) $, with $B=0.05\,$T, and a variable orientation. 

The few-particle (three-hole) eigenstates are computed by diagonalizing the interacting Hamiltonian, within the full configuration interaction approach \cite{CI,Secchi21b}. The three-hole Slater determinants are built from a set of 64 single-particle states, which ensures the convergence of the relevant quantities. Further details on the method are provided in Appendix \ref{app: Hamiltonian}. 

A final note on the choice of parameters that define the structure, and specifically the quantum well in the vertical ($z$) direction. The values we adopt for the well thickness $L$ and for the band offset $W$ are typical of MOS structures. Strained Ge wells in SiGe barriers are characterized by smaller band offsets and typically --- though not necessarily \cite{Tosato22a} --- by larger well widths. The choice of keeping also for the strained systems the same values of $L$ and $W$ is primarily motivated by the intention of comparing strained and unstrained systems on an equal footing.
Besides, as detailed in the following, the low-energy excitations in the considered systems are mainly related to the in-plane motion of the heavy-holes, which is only weakly affected by moderate changes in the vertical confinement. From a computational point of view, the current values of the well width and of the energy barrier allow us to reduce the number of basis states along the $z$ direction, and thus to expand the basis set in the planar directions. This, in turn, enables an accurate investigation of the many-body effects and of their impact on the relevant figures of merit, which is the central subject of this paper. Further investigations will be needed in order to provide a detailed comparison between different heterostructures and device designs.

\section{Single- and three-hole spin qubits}
The magnetic field breaks the Kramers degeneracy within the ground doublet, and induces the energy splitting $\Delta^{(N)}_{10}$ between
the states $\left|0^{(N)}\right\rangle$ and $\left|1^{(N)}\right\rangle$, which define the SHQ ($N=1$) and the THQ ($N=3)$.
This splitting can equivalently be expressed in terms of the $g$ factor $g^{(N)}_\theta = \Delta^{(N)}_{10} / \mu_{\rm B} B $, which depends on the orientation of $\boldsymbol{B}$ with respect to the $z$ axis, defined by $\theta$ and $\phi$; in the following, we fix $\phi = 45 \degree$.  

\begin{figure}[t]
    \centering
    \includegraphics[width=0.48\textwidth]{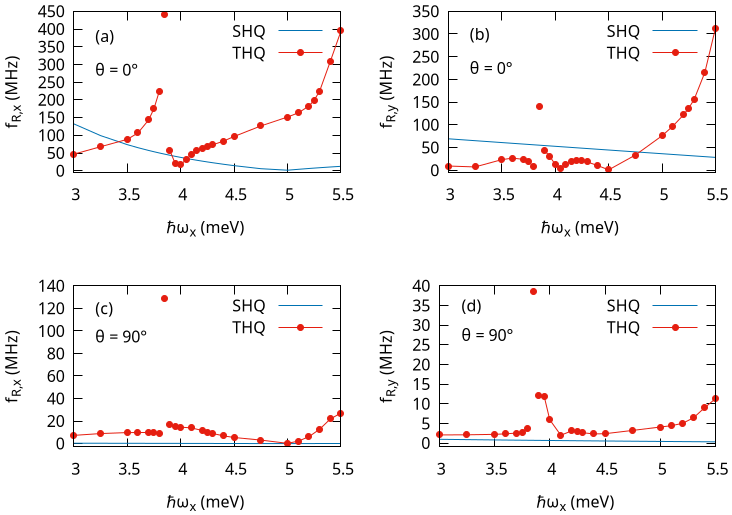} 
    \caption{Rabi frequencies $f^{(N)}_{{\rm R}, \alpha}$ of the unstrained $N$-hole qubit as functions of $\hbar \omega_x$, with $N=1$ (blue curves) and $N=3$ (red symbols), when the oscillating electric field (of amplitude $\left| \delta \boldsymbol{E}_{\rm R} \right| = 1$ mV$/$nm) is oriented along the $x$ (a, c) or $y$ direction (b, d). The magnetic field amplitude is $| \boldsymbol{B} | = 0.05$ T; its direction is given by $\theta = 0 \degree$ (a, b) or $\theta = 90 \degree$ (c, d), with $\phi = 45 \degree$. }
    \label{fig2}
\end{figure}

We start by considering the case of unstrained QDs. As a first means to characterize the three-hole ground state $|0^{(3)}\rangle$, we compute the particle density 
\begin{align}
  n(\boldsymbol{r}) = \sum_\beta n_\beta(\boldsymbol{r}) = \sum_\beta \langle 0^{(3)} | \hat{\psi}_\beta^\dagger (\boldsymbol{r})\,\hat\psi_\beta(\boldsymbol{r})| 0^{(3)} \rangle \,,
\end{align}
where $\beta$ runs over the six hole bands and the field operator $\hat{\psi}_\beta (\boldsymbol{r})$ annihilates a hole belonging to the band $\beta$ at the position $\boldsymbol{r}$. Being the particle density along the vertical direction weakly correlated with that in the planar directions (see Appendix \ref{app: Particle densities}), we focus on the in-plane particle distribution, represented by $N_z(x,y)\equiv\int dz\,n(\boldsymbol{r})$ (Fig. \ref{fig1}). The three holes clearly display a linear arrangement along the weak-confinement direction ($x$), characterized by the presence of three peaks [panel (a)], which become increasingly resolved for decreasing values of $\omega_x$ (see Appendix \ref{app: Particle densities}). Besides, the minority [$M=+3/2$, panel(b)] and majority [$M=-3/2$, panel(c)] HH components of $N_z(x,y)\equiv\int dz\,n(\boldsymbol{r})$ are clearly related to the central and external peaks, respectively: the ground state of the THQ is thus characterized by an "antiferromagnetic" ordering in the pseudospin defined by the HH components ($M=\pm 3/2$). The in-plane distribution of the LH components [panel (d)], whose weight represent less than 1\% of the total, display markedly different features.

The comparison between single- and multi-hole qubit encodings is mainly defined in terms of the respective $g$ factors and Rabi frequencies. Concerning the former ones, no significant difference is observed between SHQs and THQs. Both $g^{(1)}$ and $g^{(3)}$ are characterized by a weak dependence on the in-plane aspect ratio $\omega_x/\omega_y$ of the dot, and by a dependence on $\theta$ that, for $|\theta-90^\circ|\gg 1^\circ$, is given by: $g^{(N)}_\theta \approx g^{(N)}_{0^\circ} \, \cos\theta$. This relation results from the strong $g$ factor anisotropy ($g^{(N)}_{0^\circ} \gg g^{(N)}_{90^\circ}$), which --- in the case of the THQ --- is significantly enhanced by the Coulomb interactions (see the absolute values and the comparison with the non-interacting three-hole states for the elongated and quasi-circular QDs in Table \ref{TableI}). 

\begin{table}[b]
\centering
\renewcommand{\arraystretch}{1.5}
\begin{tabular}{|c|ccc|ccc|}
\hline
QD1 & $ f^{(1)}_{{\rm R},x}$ & $ f^{(1)}_{{\rm R},y}$ & $ \tau^{(1)}$ & $ f^{(3)}_{{\rm R},x}$ & $ f^{(3)}_{{\rm R},y}$ & $ \tau^{(3)}$ \\
\hline
$\theta=0^\circ$ & 57.5 & 27.0 & 5.78 & 23.0 & 5.30 & 5.92 \\
$\theta=90^\circ$ & 0.276 & 0.0804 & 1480 & 0.0941 & 0.157 &  1130 \\
\hline
\hline
QD2 & $ f^{(1)}_{{\rm R},x}$ & $ f^{(1)}_{{\rm R},y}$ & $ \tau^{(1)}$ & $ f^{(3)}_{{\rm R},x}$ & $ f^{(3)}_{R,y}$& $ \tau^{(3)}$ \\
\hline
$\theta=0^\circ$ & 4.48 & 10.7 & 5.93 & 63.7 & 33.9 & 4.50 \\
$\theta=90^\circ$ & 0.0223 & 0.0494 & 752 & 0.301 & 0.349 & 3690 \\
\hline
\end{tabular}
\renewcommand{\arraystretch}{1}
    \caption{Values of the Rabi frequencies (expressed in MHz) and of the dephasing time scales (in $\mu$s) of the strained single- and three-hole qubits, for $\hbar \omega_y = 6\,$meV, $B = 0.05$ T, $\left| \delta \boldsymbol{E}_{\rm R} \right| = 1$ mV/nm, and $|\boldsymbol{E}_{\rm cn}|=10^{-2}$ mV/nm. The value of $\hbar \omega_x$ is 3 and $5.5\,$meV for QD1 and QD2, respectively. }
    \label{TableII}
\end{table}

\begin{figure}[t]
    \centering 
    \includegraphics[width=0.48\textwidth]{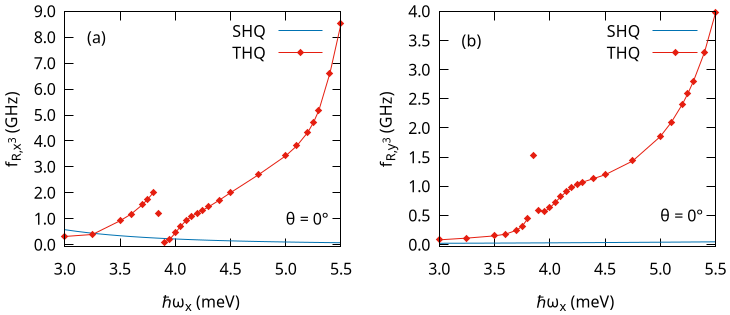}
    \caption{Rabi frequencies $f^{(N)}_{{\rm R},  \alpha}$ of the unstrained $N$-hole qubit as functions of $\hbar \omega_x$, with $N=1$ (blue curves) and $N=3$ (red symbols). The oscillating potential $V_{\rm ac}(\boldsymbol{r})$ is proportional to (a) $x^3$ or (b) $y^3$, with a proportionality constant $C=0.5885\,$meV/nm$^3$. The magnetic field amplitude is $| \boldsymbol{B} | = 0.05$ T; its direction is given by $\theta = 0 \degree$. }
    \label{fig2p}
\end{figure}

\subsection{Rabi frequencies}

The qubit manipulation is performed by applying an oscillating electric field $\delta \boldsymbol{E}(t) = \delta \boldsymbol{E}_{\rm R}\cos(\omega t) =-\nabla V_{\rm ac}(\boldsymbol{r})\cos(\omega t)$, in resonance with the Larmor frequency ($\omega = 2\pi f_{\rm L}$). The resulting Rabi frequency is given by:
\begin{align}\label{eq:rabi}
    f_{\rm R}^{(N)} = \frac{e}{h}  \Big|  \langle 0^{(N)} | V_{\rm ac}(\hat{\boldsymbol{r}})  |1^{(N)} \rangle \Big| \, ,  
\end{align}
where $\hat{\boldsymbol{r}}$ is the hole position operator. Unless differently specified, we consider hereafter the case of a spatially homogeneous AC electric field, corresponding to $V_{\rm ac} (\hat{\boldsymbol{r}}) = - \delta\boldsymbol{E}_{\rm R} \cdot \hat{\boldsymbol{r}}$, with $|\delta\boldsymbol{E}_{\rm R}|=1\,$mV/nm. 

\begin{figure}[t]
    \centering 
    \includegraphics[width=0.48\textwidth]{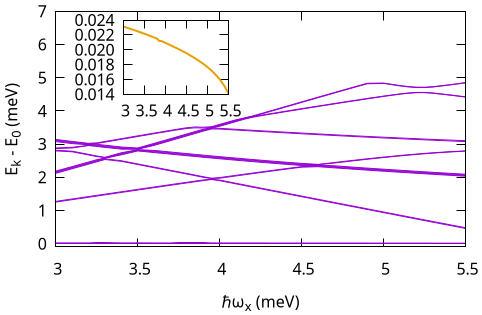}
    \caption{Excitation energies of the lowest 16 states of the unstrained three-hole system as functions of $\hbar \omega_x$, for $|\boldsymbol{B}| = 0.05$ T and $\theta = 0^{\circ}$. Figure inset: Expanded plot of the energy splitting within the ground doublet, $E_1-E_0$ (same energy units as in the main plot).}
    \label{fig5}
\end{figure}

While the $g$ factors of the THQ and of the SHQ show a very similar dependence on the orientation of $\boldsymbol{B}$, the Rabi frequencies differ remarkably in the two cases (Fig.~\ref{fig2}). In fact, for all the considered orientations of $\boldsymbol{B}$ and of the oscillating electric field $\delta \boldsymbol{E}_{\rm R}$, the Rabi frequency of the THQ displays a non-monotonic dependence on the dot aspect ratio ($\omega_x$). Moreover, the $f_{{\rm R},\alpha}^{(3)}$ are characterized by the presence of pronounced maxima, and by regions corresponding to a large enhancement with respect to the SHQ, mostly --- but not exclusively --- concentrated in the quasi-circular dot regime ($\hbar\omega_x\gtrsim 5\,$meV). The case of the circular quantum dot is not considered, because the energy gap between the ground and first-excited doublet in the THQ decreases with the dot ellipticity for $\hbar\omega_x\gtrsim 3.8\,$meV, and vanishes for $\hbar\omega_x=\hbar\omega_y=6\,$meV (see Appendix \ref{app: Levels}), making this geometry unsuitable for the implementation of the THQ.

These results are independent of the elongation direction. In fact, calculations performed with $\hbar\omega_y < \hbar\omega_x=6\,$meV give values of $f_{{\rm R},\alpha}^{(N)}$ that are specular to the ones displayed above. In other words, the Rabi frequency associated with $\delta \boldsymbol{E}_{\rm R}$ parallel to $x$, $y$, or $z$ in one case coincide, respectively, with the values of the Rabi frequency associated with $\delta \boldsymbol{E}_{\rm R}$ parallel to $y$, $x$, or $z$ in the other.

\begin{figure}[b]
    \centering 
    \includegraphics[width=0.48\textwidth]{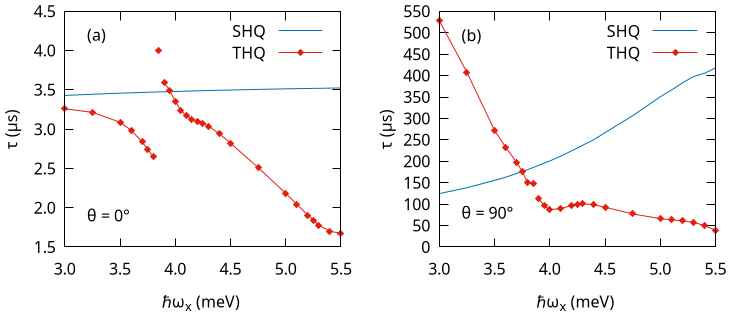}
    \caption{Dephasing time scale $\tau$ in the unstrained QDs as a function of $\hbar \omega_x$ (with $\hbar \omega_y = 6\,$meV), for a magnetic field of intensity $B = 0.05\,$T oriented (a) parallel ($\theta = 0^\circ$) or (b) perpendicular ($\theta = 90^\circ$) to the $z$ axis. The charge-noise induced electric field is oriented along the $z$ direction, with $\left| \boldsymbol{E}_{\rm cn} \right| = 10^{-2}$ mV/nm.}
    \label{fig3}
\end{figure}
Sharp peaks in the Rabi frequencies of the THQ show up at $\hbar\omega_x\approx3.8\,$meV. These features certainly are a many-body-effect, because they disappear in the non-interacting three-hole system, where the Coulomb interactions are set to zero (see Appendix \ref{app: Coulomb}). Besides, they are likely related to the presence of a narrow avoided level crossing between the first and second excited doublet, which results in a hybridization of excited three-hole states and in small but appreciable changes in the states that belong to the ground doublet (see Appendix \ref{app: Levels}).

In order to provide a wider analysis of Rabi frequencies, we have explored their dependence on spatial inhomogeneities in $\delta \boldsymbol{E}_{\rm R}$ and on the in plane-orientation of the magnetic field. In particular, concerning the first aspect, we have considered the contributions $f_{{\rm R},\alpha}^{(3)}$ resulting from nonlinear terms in the AC potential. Quadratic terms ($\alpha = x^2$ and $\alpha = y^2$) give no contribution within numerical accuracy. The cubic contributions [$\alpha = x^3,y^3$ in Fig. \ref{fig2p}(a,b)] display a dependence on $\omega_x$ that is qualitatively similar to that of the linear ones [Fig. \ref{fig2}(a,b)]. However, the ratio between the Rabi frequencies of the three- and of the single-hole qubits is much larger for the cubic than for the linear contribution, and largely exceeds $10^2$ as the circular-dot geometry is approached ($\hbar\omega_x=5.5\,$meV). For what concerns the role of the magnetic-field orientation, we find that both $f_{{\rm R},\alpha}^{(1)}$ and $f_{{\rm R},\alpha}^{(3)}$ display a cosine ($\alpha=x$) or sine ($\alpha=y$) dependence on the angle $\phi$, and therefore a cosine dependence on the relative angle between the directions of the magnetic field $\boldsymbol{B}$ and the oscillating electric field $\delta \boldsymbol{E}_{\rm R}$ (see Appendix \ref{app: Orientation}).

The question arises as to whether these striking differences between the THQ and the SHQ are mainly due to Coulomb interaction or to the occupation of excited single-particle states, imposed by the Pauli principle \cite{Fanucchi25}. In order to discriminate between these two cases, we compare the above values of the Rabi frequencies with those obtained for the noninteracting three-hole system. The difference between the two cases is always remarkable and shows that the values of the Rabi frequencies are strongly affected by  the Coulomb interactions. Whether the Rabi frequencies of the THQ are larger or smaller that those of the noninteracting three-hole system depends on the orientation of the magnetic field and on the degree of elongation of the QD (Table \ref{TableI}). 

Despite the interacting and correlated character of the three-hole states, the system is not in the Wigner-molecule (WM) regime. This is clearly seen from the plot of the excitation energies of the 3-hole system (Fig.~\ref{fig5}). There, the lowest orbital excitations (those between different Zeeman-split doublets and quadruplets) are of the order of few meV, and thus comparable to $\hbar \omega_{\alpha}$ ($\alpha=x,y$) and to the single-hole excitation energies. In the Wigner-molecule regime, instead, 3-hole excitation energies would be much smaller \cite{Secchi10}. Consistently, the plots of the charge density reveal that the density peaks, though identifiable, are not well resolved, in contrast to the typical situation in the WM regime \cite{Ercan21, Goldberg24}.

\subsection{Qubit dephasing} Random fluctuations in the energy gap $\Delta_{10}$, induced by charge noise, are generally regarded as the main source of decoherence for hole-spin qubits \cite{burkard2023}. The dephasing time scale $\tau$ can thus be defined as: 
\begin{align}
    \frac{1}{\tau^{(N)}}  
    \equiv \frac{e}{h} \Big| \boldsymbol{E}_{\rm cn} \cdot \left( \langle 1^{(N)} | \hat{\boldsymbol{r}} | 1^{(N)}\rangle - \langle 0^{(N)} | \hat{\boldsymbol{r}} | 0^{(N)} \rangle \right) \Big| \,.
    \label{T2star}
\end{align}
Here, $\boldsymbol{E}_{\rm cn}$ is a characteristic value of the charge-noise induced electric field vector. If the field is spatially homogeneous throughout the QD region then, due to the symmetries of the confinement potential, the dominant contribution in $1/\tau$ is the one related to the $z$ component of $\boldsymbol{E}_{\rm cn}$. As in the case of the Rabi frequencies, attention should be paid to the trends and to the comparison between single- and three-hole spin qubits, rather than to the absolute values of the dephasing time scale, which depend on the assumed value of the charge-noise induced electric field.

\begin{figure}[t]
    \centering 
    \includegraphics[width=0.48\textwidth]{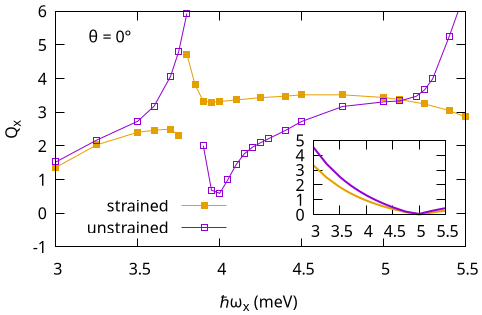} 
    \caption{Quality factor $Q_x$ for the THQ as a function of $\hbar\omega_x$ (with $\hbar\omega_y=6\,$meV), for a magnetic field intensity $B=0.05\,$T oriented out fo plane (parallel to the $z$ axis). Yellow and purple symbols correspond to the strained and to the unstrained dots, respectively. Inset: $Q_x$ {\it vs} $\hbar\omega_x$ for the SHQ, with the same units and the same color code as in the main plot.}
    \label{fig4}
\end{figure}

The differences between the values of $\tau$ obtained in the single- and in the three-hole qubits are remarkable. Similarly to the case of the Rabi frequencies, the monotonic dependence of $\tau$ on the dot aspect ratio ($\omega_x$) that is obtained for the SHQ is replaced --- in the case of the THQ --- by a non-monotonic dependence, which varies qualitatively with the magnetic-field orientation (Fig.~\ref{fig3}). In particular, we note that the dephasing time scale of the THQ is lower than that of the SHQ for a quasi-circular dot. However, the enhancement that the THQ allows in terms of Rabi frequency overcompensates such an effect, resulting in a higher value of a quality factor $ Q_x \equiv f_{{\rm R},x} \tau$ (see below). As for the Rabi frequencies, the region corresponding to $\hbar\omega_x\approx 3.8\,$meV is characterized by strong variations of $\tau$, which we attribute to the effect on the ground state doublet of the narrow anticrossing affecting the two lowest excited doublets (see Appendix \ref{app: Levels}).

\subsection{Effect of strain}
\label{subsec: Strain}
We hereafter investigate the effect of a homogeneous biaxial strain, such as the one that can result from the lattice mismatch in Ge/Si$_{1-x}$Ge$_x$ heterostructures. The Hamiltonian term related to a small and homogeneous strain is derived, within the $\boldsymbol{k} \cdot \boldsymbol{p}$ formalism, by the Bir-Pikus theory \cite{bir1974, Chao92, Secchi24}. In particular, we consider a diagonal uniaxial strain tensor, $\varepsilon_{\beta}^{\alpha}=\delta_{\beta}^{\alpha}\varepsilon_{\alpha}^{\alpha}$, 
with 
$\varepsilon_x^x=\varepsilon_y^y=\varepsilon_{\parallel}$ 
and 
$\varepsilon_z^z = -2 \varepsilon_{\parallel} C_{12} / C_{11}$, 
(where $C_{11} = 126$ GPa and $C_{12}=44$ GPa are the elastic constants of Ge \cite{Sverdlov}), and a typical Ge concentration $x = 0.8$ in the alloy, for which the measured strain parameter is $\varepsilon_{\parallel} = -0.0063$ \cite{Sammak2019}. 

The presence of strain significantly affects the values of the quantities considered so far. The strain increases the energy difference between the HH and LH components, enhancing the HH character of the single-hole ground state. This results in an increased $g$-factor anisotropy, both in the SHQ and in the THQ. As to the Rabi frequencies, their values are generally reduced by the presence of strain, to a similar extent for the SHQs and the THQs (see Appendix \ref{app: Strain}). However, the presence of biaxial strain also results in a systematic increase of $\tau^{(1)}$ and $\tau^{(3)}$, as shown in Table \ref{TableII} for the cases of the elongated and quasi-circular dots. Overall, the QD quality factor presents non-monotonic dependencies on the in-plane aspect ratio, with no general advantage or disadvantage resulting from the presence of strain (Fig.~\ref{fig4}). On the other hand, the value of $Q_x$ obtained for the THQ significantly exceeds that of the SHQ in the quasi-circular geometry and in the correspondence of the peak ($\hbar\omega_x\approx 3.8\,$meV), both for the strained and for the unstrained dot. There, the advantage of the three-hole encoding in terms of Rabi frequency ovecompensates the advantage of the SHQ at the level of dephasing time scale. On the other hand, the quality factor of the SHQ exceeds that of the THQ in the case of elongated QDs.

\section{Conclusions} The various scenarios considered in this work consistently point to the fact that three-hole spin qubits in Ge perform better than their single-hole counterparts when the QD is close to the circular regime. In both cases, the largest Rabi frequencies associated to in-plane oscillating electric fields are obtained for magnetic fields oriented along the growth direction. Biaxial strain leads to an overall reduction of the Rabi frequencies, which is however balanced by a comparable increase in the characteristic dephasing time, resulting in no significant difference between strained and unstrained dots in terms of quality factor. 

The present work only indirectly addresses some of the aspects related to the scalability of THQs. However, it was recently shown that the three-hole encoding can simultaneously suppress crosstalk and achieve high electric driving efficiencies, both of which are important ingredients for scalable architectures \cite{John25a}. As a more specific concern, three-hole systems generally present low excitation energies, which might result in unwanted state leakage and affect the qubit manipulation and readout efficiency. These effects can be limited, on the one hand, by reducing the dot symmetry and lifting the related degeneracies and, on the other hand, by avoiding highly correlated regimes, characterized by excitation energies much smaller than the single-particle ones.

\acknowledgments

We are particularly grateful to Y.-M. Niquet for an insightful comment, and we thank the members of the Niquet, Veldhorst, Rimbach-Russ, and Scappucci groups for helpful discussions. We acknowledge the CINECA award under the ISCRA initiative, for the availability of high-performance computing resources and support (IsCb4-GERONIMO - HP10CG99PB), and thank N. Spallanzani for technical support. This work was funded by the Italian PNRR Project PE0000023-NQSTI and, as a part of NCCR SPIN, a National Centre of Competence in Research, by the Swiss National Science Foundation (grant number 225153). SB was also supported by the EU through the H2024 QLSI2 project and the Army Research Office under Award Number: W911NF-23-1-0110. DL acknowledges the Deanship of Research at KFUPM and the Quantum Center for the support received under Grant no. CUP25102 and no. INQC2500, respectively. The views and conclusions contained in this document are those of the authors and should not be interpreted as representing the official policies, either expressed or implied, of the Army Research Office or the U.S. Government. The U.S. Government is authorized to reproduce and distribute reprints for Government purposes notwithstanding any copyright notation herein.

\appendix

\section{The single- and the few-hole Hamiltonians}
\label{app: Hamiltonian}

Single-hole envelope functions are obtained by diagonalizing the 6-band Luttinger-Kohn (LK) Hamiltonian \cite{Luttinger55, Voon09}, which describes the hole states in Ge close to the $\Gamma$ point. In the reference frame defined by the unit vectors $\boldsymbol{u}_x \parallel [110]$, $\boldsymbol{u}_y  \parallel [\bar{1}10]$, and $\boldsymbol{u}_z \parallel [001]$, the single-hole Hamiltonian is written in the strained-coordinate ($\boldsymbol{r}$) representation \cite{Secchi24} as:
\begin{align}
    \hat{\mathcal{H}}  \equiv \hat{\mathcal{H}}_{\boldsymbol{B} = \boldsymbol{0}} + \hat{\mathcal{H}}_{\rm ZB} + \hat{\mathcal{H}}_{\rm para} + \hat{\mathcal{H}}_{\rm dia} \,, 
    \label{total H1h}
\end{align}
where
\begin{widetext}
\begin{align}
\hat{\mathcal{H}}_{\boldsymbol{B} = \boldsymbol{0}} \equiv   \left( \begin{matrix} 
\hat{P} + \hat{Q} + V &  -\hat{S}      &  \hat{R}  &  0  &  - \frac{1}{\sqrt{2}} \hat{S}    &  \sqrt{2}  \hat{R}  \\
-\hat{S}^{\dagger}    &  \hat{P} - \hat{Q}   + V &  0   &  \hat{R}     &  - \sqrt{2} \hat{Q}     &  \sqrt{\frac{3}{2}}  \hat{S}   \\
\hat{R}^{\dagger}   &  0      &  \hat{P} - \hat{Q}   + V  &  \hat{S}   &  \sqrt{\frac{3}{2}}  \hat{S}^{\dagger}  &  \sqrt{2} \hat{Q}  \\
0	   &  \hat{R}^{\dagger}    &  \hat{S}^{\dagger}    &  \hat{P} + \hat{Q}  + V   &  - \sqrt{2}  \hat{R}^{\dagger}   &  - \frac{1}{\sqrt{2}} \hat{S}^{\dagger}  \\
      - \frac{1}{\sqrt{2}} \hat{S}^{\dagger}  &   - \sqrt{2} \hat{Q}     &  \sqrt{\frac{3}{2}}  \hat{S}  &  - \sqrt{2}  \hat{R} &  \hat{P}  + V  + \Delta_{\rm SO} & 0 \\ 
       \sqrt{2}  \hat{R}^{\dagger} & \sqrt{\frac{3}{2}}  \hat{S}^{\dagger}  &  \sqrt{2} \hat{Q}^\dagger & - \frac{1}{\sqrt{2}} \hat{S}   & 0 &  \hat{P}  + V  + \Delta_{\rm SO} 
\end{matrix}\right)    \,,
\label{B=0 Hamiltonian}
\end{align}
\end{widetext}
is the Hamiltonian in the absence of magnetic field, where $\Delta_{\rm SO} = 290$ meV is the spin-orbit coupling (SOC) parameter, and the quantities
\begin{align}
& \hat{P} = \frac{\hbar^2}{2 m_0} \gamma_1 \left( \hat k^2_x + \hat k^2_y + \hat k^2_z \right) + P_{\varepsilon} \,, \nonumber \\
& \hat{Q} = \frac{\hbar^2}{2 m_0} \gamma_2 \left( \hat k^2_x + \hat k^2_y - 2 \hat k^2_z \right) + Q_{\varepsilon} \,, \nonumber \\
& \hat{R} = \frac{\hbar^2}{2 m_0} \sqrt{3} \left[    -   \gamma_3   \left( \hat k_{x}^2 - \hat k_{y}^2 \right)  + 2 {\rm i} \gamma_2   \hat k_x \hat k_y     \right] + R_{\varepsilon} \,, \nonumber \\
&\hat{S} = \frac{\hbar^2}{2 m_0} 2 \sqrt{3} \gamma_3  \left( \hat k_x - {\rm i} \hat k_y\right) \hat k_z + S_{\varepsilon} 
\end{align}
include the $\boldsymbol{k} \cdot \boldsymbol{p}$ kinetic terms and the Bir-Pikus terms accounting for homogeneous strain, i.e.,
\begin{align}
& P_{\varepsilon} = - a_v \left( \varepsilon_{x_{\rm cr}, x_{\rm cr}} + \varepsilon_{y_{\rm cr}, y_{\rm cr}} + \varepsilon_{z_{\rm cr},z_{\rm cr}}\right) \,, \nonumber \\
& Q_{\varepsilon} = - \frac{b}{2} \left( \varepsilon_{x_{\rm cr}, x_{\rm cr}} + \varepsilon_{y_{\rm cr}, y_{\rm cr}} -2 \varepsilon_{z_{\rm cr},z_{\rm cr}}\right) \,, \nonumber \\
& R_{\varepsilon} =  {\rm i}   \frac{\sqrt{3}}{2} b \left( \varepsilon_{x_{\rm cr}, x_{\rm cr}} - \varepsilon_{y_{\rm cr}, y_{\rm cr}} \right) + d \varepsilon_{x_{\rm cr}, y_{\rm cr}} \,, \nonumber \\
& S_{\varepsilon} =  - {\rm e}^{{\rm i} \pi/4} d \left( \varepsilon_{z_{\rm cr}, x_{\rm cr}}   - {\rm i}   \varepsilon_{y_{\rm cr}, z_{\rm cr}} \right) \,.
\end{align}
Here $a_v$, $b$ and $d$ are the Bir--Pikus deformation potentials \cite{Chao92}, and the strain tensor is expressed in the crystal-axes reference frame for ease of comparison with the literature; the various terms have been transformed to the reference frame used here. In this work, we have considered either the case of no strain ($\varepsilon_{\mu,\nu} = 0$), or the biaxial strain that arises in a planar Ge/Si$_{1- X}$Ge$_{X}$ heterostructure, due to the epitaxial growth of the Ge channel on the top of a $(001)$ SiGe substrate \cite{Sverdlov}:
\begin{align}
    \varepsilon_{x_{\rm cr}, x_{\rm cr}} = \varepsilon_{y_{\rm cr}, y_{\rm cr}}   \equiv \varepsilon_{\parallel} \,, \quad \varepsilon_{z_{\rm cr}, z_{\rm cr}}= -2 \varepsilon_{\parallel} \frac{C_{12}}{C_{11}} \,,
\end{align}
where $C_{11} = 126$ GPa and $C_{12}=44$ GPa are the elastic constants of Ge \cite{Sverdlov}. For a typical Ge concentration $X=0.8$ in the alloy, which we assume in this work, one has $\varepsilon_{\parallel} = -0.0063$ \cite{Sammak2019}. 

In Eq.~\eqref{B=0 Hamiltonian}, the operator $V \equiv V_{{\rm conf};\varepsilon}(\boldsymbol{r})$ is the confining potential in the strained coordinates \cite{Secchi24}; it is assumed to be separable and given by 
\begin{align}
    V_{{\rm conf};0}(\boldsymbol{r}) = \frac{1}{2} \frac{m_0}{\gamma_1} \left( \omega_x^2 x^2 + \omega_y^2 y^2 \right)  + V_{\perp; 0}(z) 
\end{align}
in the unstrained system, where
\begin{align}
    V_{\perp; 0}(z) & = W + \frac{W}{2}\sum_{\xi=\pm 1}\tanh\left[\left(\xi z -  \frac{(1+ \xi) L_0}{2}  \right)\Big/s\right] \nonumber \\
    & \quad +   e_{\perp; 0}   z
\end{align}
is the confinement along the growth direction, which is practically a triangular well; in the chosen reference frame (where the $z$ axis points from the top gate towards the underlying heterostructure), the potential generated by the top gate attracts the holes towards $z = 0$, i.e. the upper edge of the channel. The well width is $L_0 = 10$ nm, and the electrostatic bias is $e_{\perp; 0} = 1$ meV$/$nm; we assume a barrier height $W = 4$ eV, and we use $s= 0.02$ nm as a smoothing factor for numerical convenience. In the presence of strain, as discussed in Ref.~\cite{Secchi24}, we must use the potential
\begin{align}
    V_{\varepsilon}(\boldsymbol{r}) = V_0[\boldsymbol{r} + \boldsymbol{u}(\boldsymbol{r})] = V_0(\boldsymbol{r} + \boldsymbol{\varepsilon}\boldsymbol{r}) \,.
    \label{strained potential}
\end{align}
In the case of homogeneous diagonal strain that we are considering here, the transformed potential is similar to the unstrained case, but with some small renormalizations of the parameters. More complicated couplings between the different directions would exist if the potential were not separable and/or the strain tensor were not diagonal.

Finally, the remaining terms of Eq.~\eqref{total H1h} are linear ($\hat{\mathcal{H}}_{\rm ZB}$, $\hat{\mathcal{H}}_{\rm para}$) and quadratic ($\hat{\mathcal{H}}_{\rm dia}$) in the magnetic field. In particular, $\hat{\mathcal{H}}_{\rm ZB}$ is the Zeeman-Bloch term, while the paramagnetic ($\hat{\mathcal{H}}_{\rm para}$) and diamagnetic ($\hat{\mathcal{H}}_{\rm dia}$) terms are the $\boldsymbol{B}$-dependent terms obtained from the Peierls substitution in $\hat{\mathcal{H}}_{\boldsymbol{B} = \boldsymbol{0}}$, which is done using the vector potential in the symmetric Coulomb gauge,  $\boldsymbol{A} = - \frac{1}{2} \boldsymbol{r} \times \boldsymbol{B}$.

To compute the 3-hole states within the configuration-interaction (CI) scheme, we need to evaluate the two-body matrix elements of the Coulomb interaction between single-particle states \cite{Secchi21b}. Keeping only intraband contributions, and neglecting the small corrections due to the use of strained coordinates, the quantities to be evaluated preliminarily to the CI calculation are:
\begin{align}
    & V_{\lbrace \nu_1, \nu_2, \nu_3, \nu_4 \rbrace} \nonumber \\
    & = \sum_{\tau, \sigma} \sum_{\tau', \sigma'} \int d \boldsymbol{r} \int d \boldsymbol{r}' \psi^*_{\nu_1; (\tau, \sigma)}(\boldsymbol{r}) \, \psi^*_{\nu_2; (\tau', \sigma')}(\boldsymbol{r}') \nonumber \\
    & \quad \times V_{\rm C}(\boldsymbol{r} - \boldsymbol{r}') \psi_{\nu_3; (\tau', \sigma')}(\boldsymbol{r}')  \psi_{\nu_4; (\tau, \sigma)}(\boldsymbol{r})    \,,
    \label{3D Coulomb integral}
\end{align}
where
\begin{align}
    V_{\rm C}(\boldsymbol{r} - \boldsymbol{r}') = \frac{e^2}{\epsilon \sqrt{(x-x')^2 + (y-y')^2 +  (z-z')^2 } }    
\end{align}
is the standard Coulomb potential, $\epsilon$ being the dielectric constant.

\begin{figure}[t]
\includegraphics[width=0.45\textwidth]{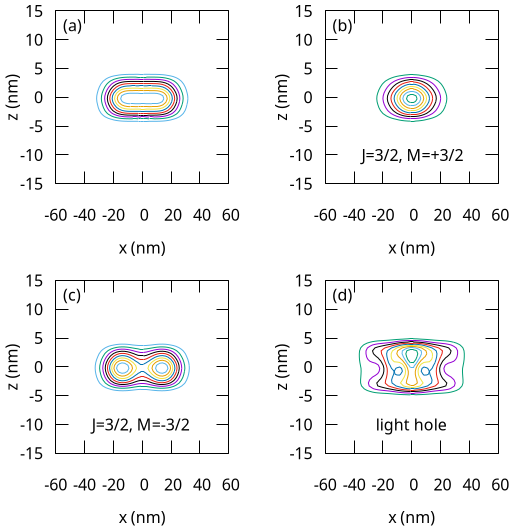}
\centering 
\caption{In-plane particle densities obtained from the ground state $|\Psi_0\rangle$ of the three-hole system. The four panels display: (a) the overall density $N_y(x,z)$; (b,c) the heavy-hole contributions; (d) the sum of the light-hole contributions. All the plotted functions are normalized to 1, the isolines correspond to values that are multiples of 0.1.  The values of the parameters are: $\hbar\omega_x=5\,$meV, $\hbar\omega_y=6\,$meV, $B=0.05\,$T and $\theta=0$ (parallel field orientation).} 
\label{figS02}
\end{figure}

\begin{figure}[t]
\includegraphics[width=0.45\textwidth]{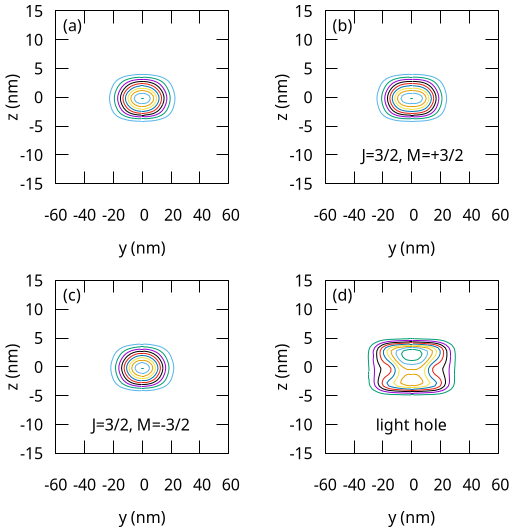}
\centering 
\caption{In-plane particle densities obtained from the ground state $|\Psi_0\rangle$ of the three-hole system. The four panels display: (a) the overall density $N_x(y,z)$; (b,c) the heavy-hole contributions; (d) the sum of the light-hole contributions. All the plotted functions are normalized to 1, the isolines correspond to values that are multiples of 0.1.  The values of the parameters are: $\hbar\omega_x=5\,$meV, $\hbar\omega_y=6\,$meV, $B=0.05\,$T and $\theta=0$ (parallel field orientation).} 
\label{figS03}
\end{figure}

\begin{figure}[t]
\includegraphics[width=0.45\textwidth]{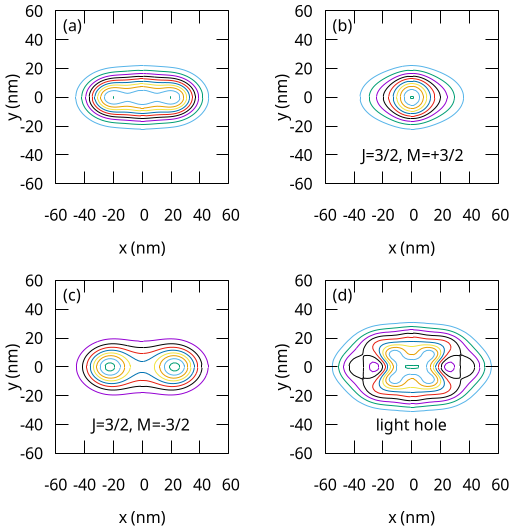}
\centering 
\caption{In-plane particle densities obtained from the ground state $|\Psi_0\rangle$ of the three-hole system. The four panels display: (a) the overall density $N_z(x,y)$; (b,c) the heavy-hole contributions; (d) the sum of the light-hole contributions. All the plotted functions are normalized to 1, the isolines correspond to values that are multiples of 0.1.  The values of the parameters are: $\hbar\omega_x=3\,$meV, $\hbar\omega_y=6\,$meV, $B=0.05\,$T and $\theta=0$ (parallel field orientation).} 
\label{figS01}
\end{figure}

\begin{figure}[t]
\includegraphics[width=0.45\textwidth]{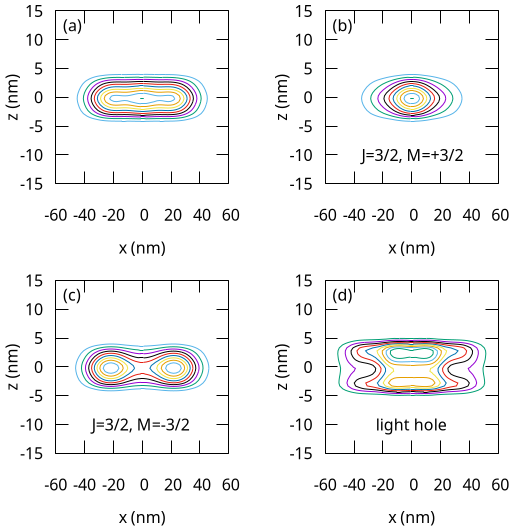}
\centering 
\caption{In-plane particle densities obtained from the ground state $|\Psi_0\rangle$ of the three-hole system. The four panels display: (a) the overall density $N_y(x,z)$; (b,c) the heavy-hole contributions; (d) the sum of the light-hole contributions. All the plotted functions are normalized to 1, the isolines correspond to values that are multiples of 0.1.  The values of the parameters are: $\hbar\omega_x=3\,$meV, $\hbar\omega_y=6\,$meV, $B=0.05\,$T and $\theta=0$ (parallel field orientation).} 
\label{figS04}
\end{figure}

\begin{figure}[t]
\includegraphics[width=0.45\textwidth]{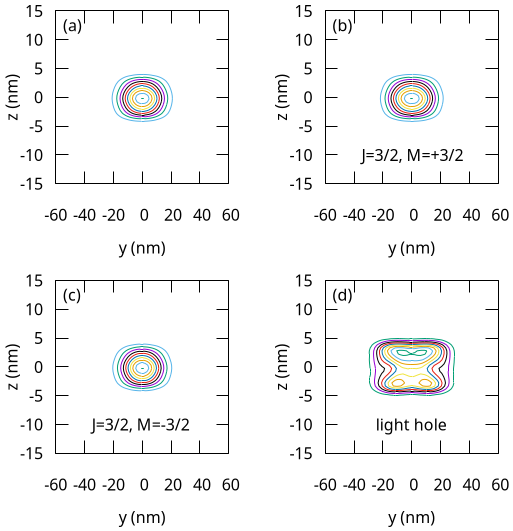}
\centering 
\caption{In-plane particle densities obtained from the ground state $|\Psi_0\rangle$ of the three-hole system. The four panels display: (a) the overall density $N_x(y,z)$; (b,c) the heavy-hole contributions; (d) the sum of the light-hole contributions. All the plotted functions are normalized to 1, the isolines correspond to values that are multiples of 0.1.  The values of the parameters are: $\hbar\omega_x=3\,$meV, $\hbar\omega_y=6\,$meV, $B=0.05\,$T and $\theta=0$ (parallel field orientation).} 
\label{figS05}
\end{figure}

\section{Particle densities}
\label{app: Particle densities}

We start by reporting the expressions that have been implemented for the calculation of the hole density.
The generic Fock state $\alpha$ for $N$ holes reads:
\begin{align}
|\Phi_\alpha\rangle = |\xi_{1,\alpha},\dots ,\xi_{N,\alpha}\rangle = c^\dagger_{\xi_{1,\alpha}} \dots c^\dagger_{\xi_{N,\alpha}} |0\rangle ,
\end{align}
where $\xi_{1,\alpha} < \xi_{2,\alpha} < \dots < \xi_{N,\alpha}$ denote spin-orbitals, such as the multiband single-hole eigenstates. The relation between the operator that annihilates a hole in the generic single-particle eigenstate $|\xi\rangle$ and the one that annihilates a hole in a generic basis set $|\lambda,\beta\rangle$ is given by the equations:
\begin{align}
& \hat c_\xi = \sum_\beta\sum_\lambda \langle \xi | \lambda,\beta\rangle \,\hat c_{\lambda,\beta} = \sum_\beta\sum_\lambda a_{\lambda,\beta;\xi}^* \,\hat c_{\lambda,\beta} \,, \nonumber \\
& \hat c_{\lambda,\beta} = \sum_\xi \langle \lambda,\beta|\xi\rangle \,\hat c_{\xi} = \sum_\beta\sum_\lambda a_{\lambda,\beta;\xi} \,\hat c_{\xi} \,,
\end{align}
where 
$|\xi\rangle = \sum_\beta\sum_\lambda a_{\lambda,\beta;\xi} |\lambda,\beta\rangle$
and $\beta\equiv (J,M)$ runs over the six (heavy-hole, light-hole, and split-off) bands.

The band-resolved fermionic field operator reads:
\begin{align}
\hat\psi_\beta(\boldsymbol{r}) & = \sum_\lambda \phi_{\lambda,\beta} (\boldsymbol{r})\, \hat c_{\lambda,\beta}
= \sum_\xi \langle \beta,\boldsymbol{r}|\xi\rangle \,\hat c_{\xi} \nonumber \\
& \equiv \sum_\xi F_\xi (\beta,\boldsymbol{r})\,\hat c_{\xi}\,,
\end{align}
where $\lambda$ specifies the orbital state (here assumed to be band dependent).
As a result, the ``band-diagonal'' one-particle density operator reads:
\begin{align}
& \hat\rho(\boldsymbol{r}) \nonumber \\
& = \sum_\beta \hat\psi_\beta^\dagger(\boldsymbol{r}) \, \hat\psi_\beta(\boldsymbol{r})   = 
\sum_{\lambda,\lambda'} \sum_{\beta}\phi_{\lambda,\beta}^* (\boldsymbol{r}) \, \phi_{\lambda',\beta} (\boldsymbol{r})  \, \hat c_{\lambda,\beta}^\dagger\,\hat c_{\lambda',\beta} \nonumber\\
& = \sum_{\xi,\xi'} \sum_\beta F_\xi^* (\beta,\boldsymbol{r}) \,F_{\xi'} (\beta,\boldsymbol{r}) \, \hat c_{\xi}^\dagger\,\hat c_{\xi'} \equiv
\sum_{\xi,\xi'} f_{\xi,\xi'} (\boldsymbol{r}) \, \hat c_{\xi}^\dagger\,\hat c_{\xi'} .
\end{align}

The $k$-th $N$-particle eigenstate can be expanded in the basis of the Fock states:
\begin{gather}
    |\Psi_k\rangle = \sum_{\alpha} C_{k,\alpha} \, |\Phi_\alpha\rangle .
\end{gather}
The expectation value of $\hat\rho(\boldsymbol{r})$ in $|\Psi_k\rangle$, the one-particle density, thus reads:
\begin{align}
& n(\boldsymbol{r}) = \langle \Psi_k | \hat\rho(\boldsymbol{r}) |\Psi_k\rangle = \sum_{\beta} n_{\beta}(\boldsymbol{r}) \nonumber \\
& = \sum_{\alpha,\beta} C^*_{k,\alpha} 
C_{k,\beta} \langle \Phi_\alpha| \hat\rho(\boldsymbol{r}_1) | \Phi_\beta \rangle \nonumber \\ 
& = \sum_{\alpha,\beta} C^*_{k,\alpha} 
C_{k,\beta} 
\sum_{\lambda,\lambda'} \sum_{\beta} \phi_{\lambda,\beta}^* (\boldsymbol{r})  \phi_{\lambda',\beta} (\boldsymbol{r}) 
\langle \Phi_\alpha | \hat c_{\lambda,\beta}^\dagger\,\hat c_{\lambda',\beta} |\Phi_\beta\rangle \nonumber \\
& =\sum_{\alpha,\beta} C^*_{k,\alpha} 
C_{k,\beta}\sum_{\xi,\xi'} f_{\xi,\xi'} (\boldsymbol{r}) \, \langle\Phi_\alpha|\hat c_{\xi}^\dagger\,\hat c_{\xi'}|\Phi_\beta\rangle \,.
\end{align}
The matrix element $\langle \Phi_\alpha| \hat c_{\xi}^\dagger\,\hat c_{\xi'} | \Phi_\beta\rangle$ can be nonzero only if the spin-orbital $\xi$ ($\xi'$) is occupied in $|\Phi_\alpha\rangle$ ($|\Phi_\beta\rangle$), and all other occupied orbitals coincide in the two Fock states. If this is the case the above matrix element takes the values of $+1$ or $-1$, depending on the relative order of the states $\xi$ and $\xi'$ within the Fock states $|\Phi_\alpha\rangle$ and $|\Phi_\beta\rangle$, respectively.

The integrated particle density $N_z(x,y)=\int n(\boldsymbol{r})\,dz$ obtained from the three-hole ground state $|\Phi_0\rangle$ is reported in Fig.~\ref{fig1} for $\hbar\omega_x = 5\,$meV. In order to complete the picture, we report here the spatial distributions $N_y(x,z)=\int n(\boldsymbol{r})\,dy$ and $N_x(y,z)=\int n(\boldsymbol{r})\,dx$ for the same quantum dot. For the sake of clarity, each of the plotted functions is normalized to 1. In order to account for the different occupation of the bands in the three-hole ground state, we report in the text the spatial integrals $A_\beta = \int d\boldsymbol{r}\, n_\beta (\boldsymbol{r})$, whose sum over the six bands corresponds to the particle number.

The function $N_y(x,z)$ [Fig.~\ref{figS02}, panel (a)] shows a linear arrangement along the $x$ direction and a weak correlation between the spatial distributions in the $x$ and $z$ directions, resulting from the strong confinement in the vertical direction. The minority HH component [$A_{(J=3/2,M=+3/2)}=0.994$, panel (b)] is characterized by a central peak; the majority HH component [$A_{(J=3/2,M=-3/2)}=1.98$, panel (c)] displays two side peaks. The sum of the two LH components [$A_{(J=3/2,M=+1/2)}+A_{(J=3/2,M=-1/2)}=0.0249$, panel (d)] displays a more complex spatial distribution with respect to the HH components. The function $N_x(y,z)$ [Fig.~\ref{figS03}, panel (a)] also shows a weak correlation between the spatial distributions in the $y$ and $z$ directions, resulting from the strong confinement in the vertical direction. The minority HH component [panel (b)] is slightly more delocalized than the majority HH component [panel (c)] along the $y$. The sum of the two LH components [panel (d)] displays also in this plane a more complex spatial distribution with respect to the HH components. Overall, the occupation of the HH states represents more than the 99\% of the overall band occupation. The picture that emerges from the above plots is thus the following: the holes in the ground state of the three-hole system display an antiferromagnetic ordering in the pseudospin $|\!\uparrow\rangle\equiv |J=3/2,M=+3/2\rangle$, $|\!\downarrow\rangle\equiv |J=3/2,M=-3/2\rangle$.

The above behaviors do not change qualitatively in a more elongated QD. To show this, in the following we plot the integrated particle densities for the case $\hbar\omega_x=3\,$meV. The function $N_z(x,y)$ [Fig.~\ref{figS01}, panel (a)] shows a linear arrangement along the $x$ direction, characterized by the presence of three peaks. The central peak is related to the minority HH component [$A_{(J=3/2,M=+3/2)}=0.995$, panel (b)]; the two side peaks are related to the majority HH component [$A_{(J=3/2,M=-3/2)}=1.99$, panel (c)]. The sum of the two LH components [$A_{(J=3/2,M=+1/2)}+A_{(J=3/2,M=-1/2)}=0.0249$, panel (d)] displays a more complex spatial distribution with respect to the HH components. This linear arrangement, essentially characterized by an alternation of $M=-3/2$ and $M=+3/2$ HH components, also emerges from the plots of the $N_y(x,z)$ (Fig.~\ref{figS04}) and $N_x(y,z)$ (Fig.~\ref{figS05}) functions, qualitatively similar to the ones obtained for the quasi-circular QD ($\hbar\omega_x=5\,$meV).

\section{Avoided level crossing}
\label{app: Levels}

By examining the plot of the three-hole excitation energies (Fig.~\ref{fig5}), it is seen that a level crossing occurs between excited states, exactly in the region where peaks or sharp transitions are obtained for the three-hole Rabi frequencies $f_{\rm R}^{(3)}$ and for the dephasing time scale $\tau$ (see Figs.~\ref{fig2}, \ref{fig2p}, \ref{fig3}). The fact that this is an avoided crossing, and not a true crossing, results from the inspection of the three-hole wave functions corresponding to the eigenstates $|\Psi_k \rangle$ belonging to the first and second excited doublets ($k=2,3,4,5$). These eigenstates undergo a mixing for $3.8\,$meV$ \lesssim\hbar\omega_x\lesssim 3.9\,$meV, where the gap between the two excited doublets displays a minimum. 

The same applies to the strained QD, where an avoided level crossing between the first and second excited doublets takes place in the region $3.8\,$meV$ \lesssim\hbar\omega_x\lesssim 3.9\,$meV. This coincides with the region where strong variations in the three-hole Rabi frequencies (see Appendix \ref{app: Strain} below) and quality factor (Fig.~\ref{fig4}) are obtained in the case of a strained QD.

In order to better clarify the nature of the level crossings, we compute the differential fidelity
\begin{gather}
    F_{jk} (\omega_x',\omega_x) = |\langle \Psi_j(\omega_x')|\Psi_k(\omega_x) \rangle|^2\,.
\end{gather}
If a small increase in the value of $\omega_x$ does not lead to significant changes in the three-hole eigenstate $|\Psi_j\rangle$, then $F_{jj}$ is close to 1 and all the $F_{jk}$ with $k\neq j$ are close to 0. If, instead, $F_{jj}$ is significantly lower than 1 and some of the $F_{jk}$ is nonnegligible, then a small variation in $\omega_x$ induces a mixing between the three-hole eigenstate $|\Psi_j\rangle$ and $|\Psi_k\rangle$. 

\begin{figure}[t]
\includegraphics[width=0.48\textwidth]{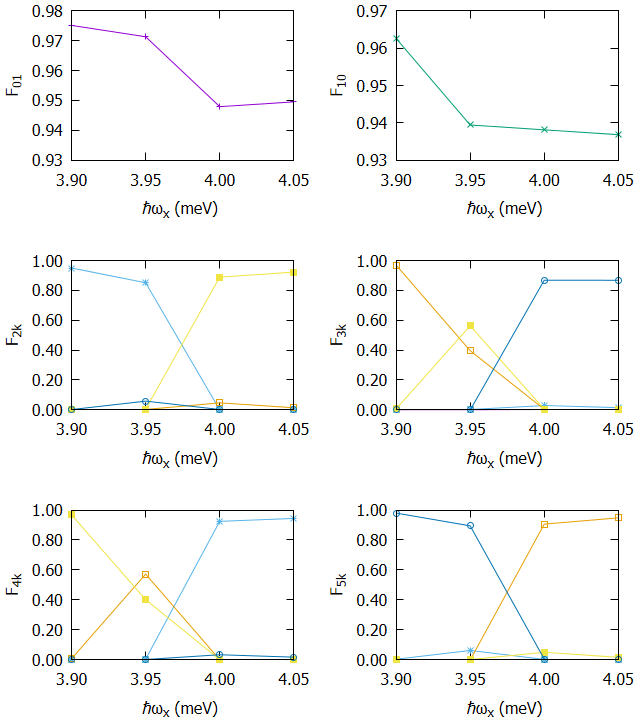}
\centering 
\caption{Differential fidelity $F_{jk}$ between three-hole eigenstates $|\Psi_j\rangle$ at $\hbar\omega_x'=3.85\,$meV and the eigenstate $|\Psi_k\rangle$ at the value of $\hbar\omega_x$ denoted in the horizontal axis. Different values of $k$ correspond to different colors in the six panels: $k=0$ is purple (symbol $+$), $k=1$ is green (symbol $\times$), $k=2$ is light blue (symbol $\ast$), $k=3$ is orange (symbol $\square$), $k=4$ is yellow (symbol $\blacksquare$), $k=5$ is blue (symbol $\bigcirc$). The values of the parameters are: $\hbar\omega_y=6\,$meV, $B=0.05\,$T and $\theta=0$ (parallel field orientation).} 
\label{figS13}
\end{figure}

From the quantities $F_{01}$ and $F_{10}$, reported in the upper panels of Fig.~\ref{figS13}, it follows that the states belonging to the ground doublet undergo a small but appreciable variation for $\hbar\omega_x\approx 3.9\,$meV, resulting from a slight mixing with excited states. A much more dramatic transition takes place in the same region with the states that belong to the first and second excited doublets (middle and lower panels, respectively). The fidelity between each of the states  $|\Psi_k\rangle$ (with $k>1$) on opposite sides of the critical region are approximately zero. In particular, the states $|\Psi_2\rangle$ (yellow) and $|\Psi_3\rangle$ (orange) evolve into $|\Psi_4\rangle$ (light blue) and  $|\Psi_5\rangle$ (blue), respectively, and viceversa. The complete mixing between the states $|\Psi_3\rangle$ and $|\Psi_4\rangle$ for $\hbar\omega_x = 3.95\,$meV (middle right and bottom left panels) demonstrates that the two states undergo an avoided level crossing, and not a crossing, as a function of $\omega_x$. It is reasonable to expect, on the basis of symmetry arguments, that the same occurs for the other  pair of states that cross, $(|\Psi_2\rangle,|\Psi_5\rangle)$, and probably also for the pairs $(|\Psi_2\rangle,|\Psi_4\rangle)$ and $(|\Psi_3\rangle,|\Psi_5\rangle)$. However, all these avoided level crossings probably are very narrow, and the resulting mixing of the diabatic states cannot be seen with the current grid of $\omega_x$ values.

\begin{figure}[t]
\includegraphics[width=0.48\textwidth]{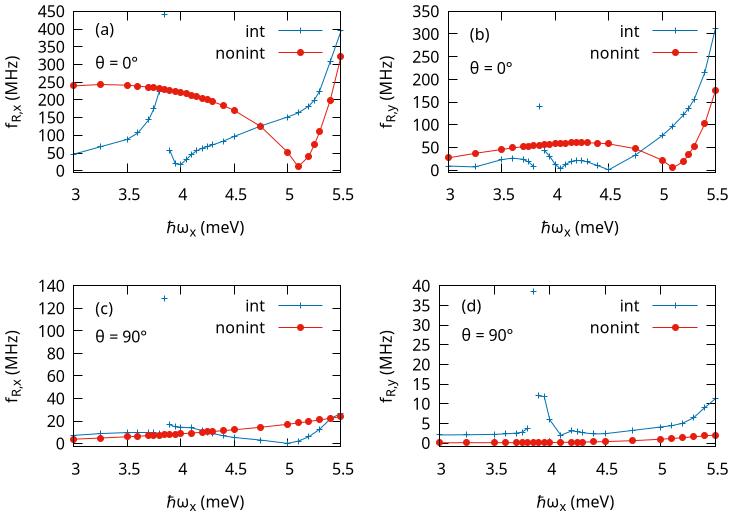}
\centering 
\caption{Rabi frequencies $f^{(3)}_{{\rm R}, \alpha}$ of the three-hole qubit as functions of $\hbar \omega_x$ in the absence of Coulomb interactions (red circles), when the oscillating electric field (of amplitude $\left| \delta \boldsymbol{E}_{\rm R} \right| = 1$ mV$/$nm) is oriented along the $x$ (a, c) or $y$ direction (b, d). The Rabi frequencies obtained in the presence of Coulomb interactions, already plotted in Fig.~\ref{fig2}, are reported here for a comparison (light blue). The magnetic field amplitude is $| \boldsymbol{B} | = 0.05$ T; its direction is given by $\theta = 0 $ (a, b) or $\theta = \pi/2$ and $\phi = \pi/4$ (c, d).} 
\label{figS06}
\end{figure}

\begin{figure}[t]
\includegraphics[width=0.3\textwidth]{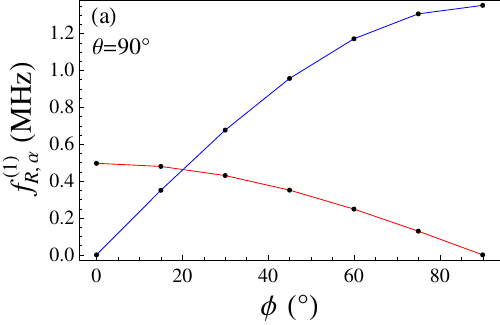}
\includegraphics[width=0.3\textwidth]{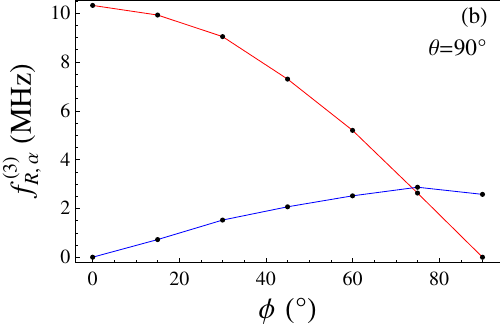}
\centering 
\caption{Dependence of the Rabi frequencies on the in-plane magnetic-field orientation for (a) the unstrained single- and (b) the three-hole qubits. Red and blue curves correspond to the cases where the electric field oscillates along the $x$ and $y$ axes, respectively. The values of the remaining parameters are: $\hbar\omega_x=3$ meV, $\hbar\omega_y=6$ meV, and $B=0.05$ T.} 
\label{figS15}
\end{figure}

\begin{figure}[t]
\includegraphics[width=0.48\textwidth]{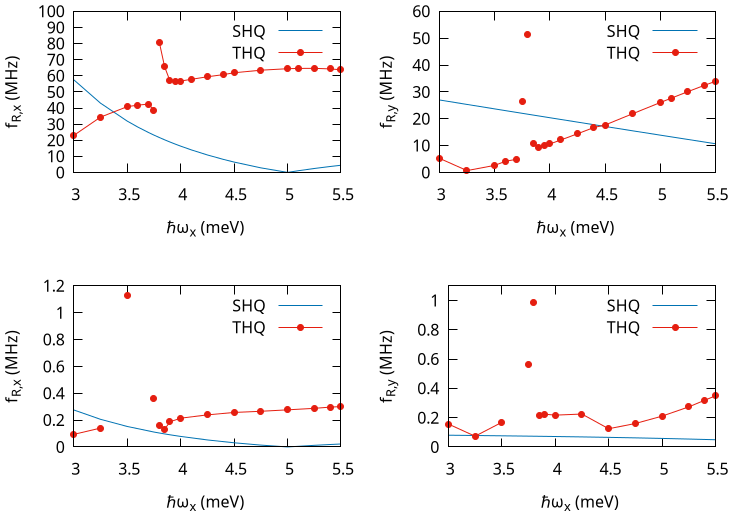}
\centering 
    \caption{Rabi frequencies $f^{(N)}_{{\rm R}, \alpha}$ of the $N$-hole qubit as functions of $\hbar \omega_x$, with $N=1$ (blue curves) and $N=3$ (red symbols), when the oscillating electric field (of amplitude $\left| \delta \boldsymbol{E}_{\rm R} \right| = 1$ mV$/$nm) is oriented along the $x$ (a, c) or $y$ direction (b, d). The calculations are performed in the presence of homogeneous biaxial strain (see Section \ref{subsec: Strain} for the strain tensor values). The magnetic field amplitude is $| \boldsymbol{B} | = 0.05$ T; its direction is given by $\theta = 0 $ (a, b) or $\theta = \pi/2$ and $\phi = \pi/4$ (c, d).}
\label{figS07}
\end{figure}

\section{Role of the Coulomb interactions}
\label{app: Coulomb}

To highlight the role played by the Coulomb interactions, we compare the values of the Rabi frequencies reported in Fig.~\ref{fig2} with those obtained for the noninteracting case. Here, the three-hole ground state corresponds to a single configuration (Slater determinant), and specifically with that where the three lowest-energy single-particle spin-orbitals are occupied. The behaviors are qualitatively different with respect to those obtained for the interacting case (Fig.~\ref{figS06}). In particular, the peak between $\hbar\omega_x=3.5\,$meV and $\hbar\omega_x=4\,$meV disappears for all the Rabi frequencies: the peak can thus be related to the Coulomb interactions and to the resulting multi-configurational character of the three-hole ground state $|\Psi_0\rangle$. 

No systematic difference emerges between the noninteracting and the interacting Rabi frequencies, i.e. neither is higher than the other one for all values of $\omega_x$. For a parallel orientation of the magnetic field [panels (a,b), $\theta =\pi/2$], the Coulomb interactions produce a reduction or an enhancement of the Rabi frequencies, depending on whether the dot is elongated ($\hbar\omega_x\lesssim 4.7\,$meV) or quasi-circular ($\hbar\omega_x\gtrsim 4.7\,$meV). However, the opposite applies to the case of a perpendicular magnetic field orientation, if the oscillating electric field is oriented along the $x$ direction [panel (c)]. The interacting Rabi frequencies are larger than the noninteracting ones in all the considered range of $\omega_x$ values, for $\theta=\pi/2$ and $\boldsymbol{E}_{\rm R} \parallel\boldsymbol{u}_x$.

\section{Role of the magnetic-field orientation}
\label{app: Orientation}

Our analysis of the dependence of the Rabi frequencies on the magnetic-field orientation has mainly focused so far on the role of the polar angle $\theta$. In order to complete this analysis, we consider here the dependence of $f_{{\rm R},\alpha}^{(N)}$ on the azimuthal angle $\phi$, for $\theta=\pi/2$ (Fig.~\ref{figS15}). Both for the single- and for the three-hole qubits, the Rabi frequencies display a cosine-like or a sine-like dependence on $\phi$ when the electric field is oriented along the $x$ or $y$ directions, respectively. This behavior can be summarized by saying that the Rabi frequencies vary as (the modulus of) a cosine function, whose argument corresponds to the difference between the in-plane orientations of the static magnetic field and of the oscillating electric field. 

The fact that both the single- and the three-hole qubits are characterized by the same angular dependence implies that the values of $\phi$ does not affect the relative figure of merit between the two. We also note that, if the magnetic field is tilted away from the $xy$ plane ($\theta < \pi/2$), the contribution to the Rabi frequencies related to the vertical ($z$) component of the field overwhelms that coming from the in-plane component, so that the dependence on the in-plane orientation of the magnetic field ($\phi$) becomes essentially irrelevant.

\section{Role of the biaxial strain}
\label{app: Strain}

We consider the effect of a homogeneous biaxial strain, such as the one that can result from the lattice mismatch in Ge/Si$_{1-x}$Ge$_x$ heterostructures. The strain tensor is defined in Sec.~\ref{subsec: Strain}. 

The Rabi frequencies corresponding to parallel and perpendicular orientations of the magnetic field, and to an oscillating electric field $\boldsymbol{E}_{\rm R}$ oriented along the $x$ and $y$ directions, are plotted in Fig.~\ref{figS07}. The effect of strain is highlighted by the comparison with the Rabi frequencies obtained for the unstrained QD, which are reported in Fig.~\ref{fig2}. Overall, the strain induces a significant reduction of the Rabi frequencies (by at least a factor 2) for all values of $\omega_x$, both for the single- and for the three-hole qubits. However, as shown by the values of the quality factor (Fig.~\ref{fig4}), this is compensated by a comparable increase of the dephasing time scale $\tau$.

\end{document}